\RequirePackage{lineno} 
\documentclass[12pt,onecolumn,prc,showpacs,preprintnumber,amsmath,
floatfix,amssymb]{revtex4-1}

\usepackage{epsfig}
\usepackage{graphicx}
\usepackage{dcolumn}
\usepackage{bm}
\usepackage{multirow}           

\def\var{\mbox{\boldmath $\varepsilon$}}
\def\p{\mbox{\boldmath $p$}}
\def\q{\mbox{\boldmath $q$}}
\def\k{\mbox{\boldmath $k$}}

\allowdisplaybreaks[4]  

\begin{document}
\title{Test of the GENIE neutrino event generator against reduced cross
  sections extracted from ${}^{16}$O$(e,e'p)$ data}
\author{A.~V.~Butkevich and S.~V.~Luchuk}
\affiliation{Institute for Nuclear Research,
Russian Academy of Sciences, Moscow 117312, Russia\\}
\date{\today}

\begin{abstract}

The reduced cross section of the semiexclusive $(l,l'p)$ lepton scattering
process irrespective of the type of interaction is determined  mainly by the
bound nucleon momentum distribution in the target and nucleon final state
interaction with the residual nucleus. These cross sections can be identified
with distorted nuclear spectral functions and therefore, they are similar up to
Coulomb corrections for neutrino and electron scattering on nuclei. In this
article, we exploit this similarity and use data with precise kinematics and
large statistics for semiexclusive electron scattering on an oxygen target to
test models employed in the GENIE neutrino event generator. We find that these
models cannot reproduce well the measured reduced cross sections in all
allowed kinematic regions and the GENIE event generator needs to describe
better both the nuclear ground states and nucleon final state interaction. The
approach presented in this paper provides a great opportunity to better test
the accuracy of nuclear models of quasielastic neutrino-nucleus scattering,
employed in neutrino event generators. 
  
\end{abstract}
 \pacs{25.30.-c, 25.30.Bf, 25.30.Pt, 13.15.+g}

\maketitle

\section{Introduction}

In current~\cite{NOvA1, T2K} and future~\cite{ DUNE,HK2T, SBN}
accelerator-based neutrino experiments the probabilities of neutrino
oscillations as functions of neutrino energy are measured to evaluate the
oscillation parameters and to test the three-flavor paradigm. The accuracy to
which neutrino oscillation parameters can be extracted depends on the ability of
experiments to determine the individual energy of each detected neutrino. These
experiments rely on neutrino event generator codes, which are simulations of
neutrino-nucleus interactions used to estimate efficiencies, backgrounds, and
systematic uncertainties for the obtained result.

The neutrino beams have broad distributions that range from tens of MeVs to a
few GeVs. The incident neutrino energy is not known a priori and can only be
deduced by measuring the leptonic and hadronic energy of the final state
particles. The energy and scattering angle of the outgoing muon can be directly
measured. To determine the hadronic energy $\var_h$ a calorimetric method is
used, which relies not only on the visible energy $E_{vis}$ measured in the
detector, but on a fit of $\var_h$ as a function of $E_{vis}$, obtained
from simulation.  Thus, the accuracy of the neutrino energy reconstruction
depends on the models of the neutrino-nucleus interactions that are implemented
 in neutrino event generators. The systematic uncertainties of this procedure 
affect significantly the determination of the incident neutrino energy.
For instance, the estimated muon and hadronic energy resolutions are 3.5\% and
25\%, respectively, giving an overall energy resolution for neutrino events in
the NOvA detector~\cite{NOvA2} of about 7\% . 

In addition, the neutrino-nucleus scattering model is critical for
obtaining background estimates in analyses aimed at determining the neutrino
oscillation parameters. Further progress in reducing systematic errors in
neutrino oscillation experiments requires a more extensive use of measurements
of final-state protons. It is needed for neutrino energy reconstruction
because a more exclusive final-state measurement allows us to better estimate
the neutrino energy and provide information about the multinucleon contribution
to the inclusive cross section. Inclusive reactions
are relatively insensitive to the details of the final nuclear states.

Neutrino- and electron-nucleus interactions are similar due to their shared
origin in electroweak theory. Electrons interact via a vector current, and
neutrinos interact via vector and axial-vector currents. The nuclear structure
and final state interactions of the outgoing hadrons they are expected to be
identical for both leptons. These similarities can be used to test models of
neutrino-nucleus interaction.
Neutrino event generator GENIE~\cite{Ankow, e4v1, CLAS, Ruso},
has been extended to electron scattering, and their results were compared with
measured inclusive cross sections of electron-nucleus ($eA$) interaction.
Note that inclusive reactions are relatively insensitive to the details of the
final nuclear states.
Therefore, rather simple phenomenological and factorized models and event
generators that employ them are able to describe inclusive and total cross
sections but cannot make predictions on both leptons and hadrons in the final
states. Theoretical models attempting to describe the semiexclusive $(l,l'p)$
lepton scattering process must be macroscopic and unfactorized models, and
simultaneously account for various nuclear effects, including nuclear ground
states, i.e. shall structure of nuclei, nucleon-nucleon (NN) correlations and
final state interaction (FSI) of the knocked-out nucleon with the residual
nucleus. 
The semiexclusive $(l,l'p)$ lepton scattering process involves the specific
asymptotic states and allows us to test the nuclear in model more in detail.
The reduced cross section, obtained from the measured differential
semiexclusive electron scattering cross section divided by the kinematic
factor and the off-shell electron-proton cross section, can be identified as
the distorted spectral function. Thus, irrespective of the type of interaction
(electromagnetic or weak), the distorted spectral function is determined mainly
by the intrinsic properties of the target and the ejected nucleon's interaction
with the residual nucleus and depends upon the initial and ejectile nucleons'
momenta and the angle between them.
Microscopic and unfactorized the relativistic distorted wave impulse
approximation (RDWIA), initially designed for the description of exclusive
$(e,e'p)$ data~\cite{Pick, Udias, JKelly} was used
~\cite{BAV1, BAV2, BAV3, BAV4} for the calculation of the charged-current (CC)
quasielastic (QE) neutrino scattering reduced cross sections as a function of
the missing nucleon momentum. The results were compared with those obtained from
measurements of $(e,e^{\prime}p)$ scattering on oxygen, carbon, calcium, and
argon targets. It was shown that these cross sections are similar to those of
electron scattering data and agree with electron scattering data.
This approach provides novel constraints on nuclear
models of the CCQE scattering and can be applied to test spectral functions and
FSI employed in neutrino event generators.
A systematic comparison of the CCQE reduced cross sections calculated within
the models employed in the GENIE event
generator~\cite{GEN1, GEN2} with those measured in electron scattering on
carbon targets was carried out in Ref.~\cite{BAV5}. 
The goal of the present paper is to test the GENIE models against 
semiexclusive electron scattering data on oxygen. This target is the main
detector component in water Cherenkov neutrino detectors, and for oxygen, there
are high-precision data sets with different monochromatic electron beams.

The outline of this article is as follows. In Sec. II we briefly present the
formalism for the CCQE semiexclusive scattering process, the calculation of
the reduced cross sections with a neutrino event generator, and regard basic
aspects of comprehensive model configurations employed in the GENIE version 3
simulation framework. Results are presented and discussed in Sec. III.
The conclusions are summarized in Sec. IV.
 
\section{Formalism of quasielastic scattering and neutrino event generator as
a tool in reduced cross section study}

\subsection{CCQE lepton-nucleus cross sections}

In the laboratory frame, the differential cross section for exclusive
electron ($\sigma ^{el}$) and (anti)neutrino ($\sigma ^{cc}$) CC scattering can
be written as
\begin{subequations}
\begin{align}
\label{1} 
\frac{d^6\sigma^{el}}{d\varepsilon_f d\Omega_f d\varepsilon_x d\Omega_x} 
&=
\frac{\vert\p_x\vert\varepsilon_x}{(2\pi)^3}\frac{\varepsilon_f}
{\varepsilon_i}
 \frac{\alpha^2}{Q^4} L_{\mu \nu}^{(el)}\mathcal{W}^{\mu \nu (el)}
\\                                                                  
\frac{d^6\sigma^{cc}}{d\varepsilon_f d\Omega_f d\varepsilon_x d\Omega_x} 
&=
\label{2} 
\frac{\vert\p_x\vert\varepsilon_x}{(2\pi)^5}\frac{\vert\k_f\vert}
{\varepsilon_i} \frac{G^2\cos^2\theta_c}{2} L_{\mu \nu}^{(cc)}
\mathcal{W}^{\mu \nu (cc)},
\end{align}
\end{subequations}
where
$k_i=(\varepsilon_i,\k_i)$ and $k_f=(\varepsilon_f,\k_f)$ are the initial and
final lepton 4-momenta, $p_A=(\varepsilon_A,\p_A)$, and
$p_B=(\varepsilon_B,\p_B)$ are the initial and final target 4-momenta,
$p_x=(\varepsilon_x,\p_x)$ is the ejectile nucleon 4-momentum, $q=(\omega,\q)$ 
is the the 4-momentum transfer carried by the virtual photon (W-boson), and
$Q^2=-q^2=\q^2-\omega^2$ is the photon (W-boson) virtuality. In
Eqs. (\ref{1}) and (\ref{2}) $\Omega_f$ is the solid angle for the lepton
momentum, $\Omega_x$ is the
 solid angle for the ejectile nucleon momentum, $\alpha\simeq 1/137$ is the
fine-structure constant, $G \simeq$ 1.16639 $\times 10^{-11}$ MeV$^{-2}$ is
the Fermi constant, $\theta_C$ is the Cabibbo angle
($\cos \theta_C \approx$ 0.9749), $L^{\mu \nu}$ is the lepton tensor, and
 $\mathcal{W}^{(el)}_{\mu \nu}$ and $\mathcal{W}^{(cc)}_{\mu \nu}$ are
respectively the electromagnetic and weak CC nuclear tensors~\cite{BAV1}.

For exclusive reactions in which only a single discrete state or a narrow
resonance of the target is excited, it is possible to integrate over the
peak in missing energy and obtain a fivefold differential cross section of
the form
\begin{subequations}
\begin{align}
\label{cs5:el}
\frac{d^5\sigma^{el}}{d\varepsilon_f d\Omega_f d\Omega_x} &= R
\frac{\vert\p_x\vert\tilde{\varepsilon}_x}{(2\pi)^3}\frac{\varepsilon_f}
{\varepsilon_i} \frac{\alpha^2}{Q^4} L_{\mu \nu}^{(el)}W^{\mu \nu (el)}
\\                                                                       
\label{cs5:cc}
\frac{d^5\sigma^{cc}}{d\varepsilon_f d\Omega_f d\Omega_x} &= R
\frac{\vert\p_x\vert\tilde{\varepsilon}_x}{(2\pi)^5}\frac{\vert\k_f\vert}
{\varepsilon_i} \frac{G^2\cos^2\theta_c}{2} L_{\mu \nu}^{(cc)}W^{\mu \nu (cc)},
\end{align}
\end{subequations}
where $R$ is a recoil factor
  \begin{equation}\label{Rec}
R =\int d\varepsilon_x \delta(\varepsilon_x + \varepsilon_B - \omega -m_A)=
{\bigg\vert 1- \frac{\tilde{\varepsilon}_x}{\varepsilon_B}
\frac{\p_x\cdot \p_B}{\p_x\cdot \p_x}\bigg\vert}^{-1},                    
\end{equation}
in which $\tilde{\varepsilon}_x$ is the solution to the equation
$
\varepsilon_x+\varepsilon_B-m_A-\omega=0,
$
where $\varepsilon_B=\sqrt{m^2_B+\p^2_B}$, $~\p_B=\q-\p_x$ and $m_A$ and $m_B$
are masses of the target and the recoil nucleus, respectively. Note that the
missing momentum is $\p_m=\p_x-\q$ and missing energy $\var_m$ are defined by
$\var_m=m + m_B -m_A$~\cite{BAV1}.

The leptonic tensor is as written in Ref.~\cite{BAV1}.
The electromagnetic and the weak CC hadronic tensors, 
$\mathcal{W}^{(el)}_{\mu \nu}$ and $\mathcal{W}^{(cc)}_{\mu \nu}$ are given
by bilinear products of the transition matrix elements of the nuclear
electromagnetic or CC operator $J_{\mu}^{(el)(cc)}$ between the initial
nuclear state $\vert A \rangle $ and the final state $\vert B_f \rangle$ as
\begin{eqnarray}
\mathcal{W}_{\mu \nu }^{(el)(cc)} &=& \sum_f \langle B_f,p_x\vert
J^{(el)(cc)}_{\mu}\vert A\rangle \langle A\vert
J^{(el)(cc)\dagger}_{\nu}\vert B_f,p_x\rangle              
\delta (\varepsilon_A + \omega - \varepsilon_x -
\varepsilon_{B_f}),
\label{W}
\end{eqnarray}
where the sum is taken over undetected states.

 The reduced cross section is given by 
\begin{equation}
\sigma_{red} = \frac{d^5\sigma^{(el)(cc)}}{d\varepsilon_f d\Omega_f d\Omega_x}
/K^{(el)(cc)}\sigma_{lN},                                                 
\label{Sred}
\end{equation}
where
$K^{el} = R {p_x\varepsilon_x}/{(2\pi)^3}$ and
$K^{cc}=R {p_x\varepsilon_x}/{(2\pi)^5}$
are phase-space factors for electron and neutrino scattering  and $\sigma_{lN}$
is the corresponding elementary cross section for lepton scattering from the
moving free nucleon~\cite{BAV5}

 The single-nucleon charged current has the $V{-}A$
 structure $J^{\mu (cc)} = J^{\mu}_V + J^{\mu}_A$. For calculation of the vertex
 function $\Gamma^{\mu (cc)} = \Gamma^{\mu}_V + \Gamma^{\mu}_A$ of a moving but free
 nucleon we employ the CC1 de Forest prescription for the off-shell vector
 current vertex function~\cite{deForest}
\begin{equation}
\Gamma^{\mu}_V = G_M(Q^2)\gamma^{\mu} - \frac{{\bar P}^{\mu}}{2 m}F_M(Q^2)  
\end{equation}
and the axial current vertex function
\begin{equation} 
\Gamma^{\mu}_A = F_A(Q^2)\gamma^{\mu}\gamma_5 + F_P(Q^2)q^{\mu}\gamma_5,  
\end{equation}
where $\bar{P}=(\var_f + \bar{\var}, 2\p_x - \q)$ and
$ \bar{\var}=\sqrt{m^2 + (\p_x -\q)^2}$.
The weak vector form factors $F_V$ and $F_M$ are related to corresponding
electromagnetic ones for proton $F^{(el)}_{i,p}$ and neutron $F^{(el)}_{i,n}$
by the hypothesis of conserved vector current 
\begin{equation}
F_i = F^{(el)}_{i,p} - F^{(el)}_{i,n}~~~~(i=V,M).                          
\end{equation}
The experimental momentum distribution is also obtained using Eq.(\ref{Sred})
with the off-shell electron-nucleon cross section $\sigma_{cc1}$ developed by de
Forest~\cite{deForest} that are normally used for $\sigma_{eN}$.
For the axial $F_A$ and pseudoscalar $F_P$ form factors, we use the dipole
approximation.
 
The reduced cross section $\sigma_{red}$ effectively represents the distorted
spectral function where final-state interactions (FSI) introduce dependencies
on the ejectile momentum, the angle between the initial and final nucleon
momentum and the incident lepton energy. In the nonrelativistic
PWIA limit, $\sigma_{red}$ reduces to the bound-nucleon momentum distribution.
These cross sections for (anti)neutrino  scattering off
nuclei are similar to those for electron scattering apart from small
differences at low beam energy due to effects of Coulomb distortion of the
incoming electron wave function as shown in Refs.~\cite{BAV1, BAV2, BAV3}.  
With the reduced cross section the factorization approximation to the
semiexclusive cross section Eq.~\ref{cs5:el} and Eq.~\ref{cs5:cc} can be
written as~\cite{BAV5}
\begin{equation}
\frac{d^5\sigma^{(el)(cc)}}{d\varepsilon_f d\Omega_f d\Omega_x}=
K^{(el)(cc)}\times \sigma_{lN} \times  \sigma_{red}(\var_m, {\p}_m, {\p}_x).    
\end{equation}
Such factorization is not strictly valid relativistically because the binding
potential alters the relationship between lower and upper components of a
Dirac wave function ~\cite{Caballero}.

\subsection{Neutrino event generator as a tool in reduced cross section study}

Reduced cross sections calculated for the oxygen nucleus within the GENIE
version 3.4.0 framework are compared with data to test the nuclear models that
are employed in the GENIE event generator. Several models for the nuclear
ground state and several models for FSI are offered within this framework. We
simulated only CCQE neutrino interactions with the oxygen nucleus. The events
with one proton and any number of neutrons which did not undergo final state
interactions are selected as a $1\mu1p$ signal. This event selection is
performed using two methods: one is based on a specific event topology in the
final state and the other relies only on the knowledge of the kinematics of
events~\cite{BAV5}.


In the Saclay experiment Ref.~\cite{Saclay} the ${}^{16}$O$(e,e^{\prime}p)$
measurements have been
performed in the perpendicular kinematics for $-100 \le p_m \le 300$ MeV/c and
the reduced cross section has been integrated in the intervals of missing energy
$\var_m =10 - 15$ MeV and 15-20 MeV, where the $p_{1/2}$ and $p_{3/2}$ hole
strengths are mostly centered in two peaks with separation energies of 12.2
MeV and 18.5 MeV, respectively.
In perpendicular kinematics, the incident energy and energy transferred are
fixed, as are the electron scattering angle and the outgoing proton energy,
whereas the missing momentum changes with the proton angle. It is necessary to
choose the electron angle and outgoing proton energy such that $|\p_x|=|\q|$.
The vectors $\p_m$ and $\q$ are almost perpendicular.   

In the NIKHEF experiment Ref.~\cite{NIK1, NIK2, NIK3} the ${}^{16}$O reduced
cross section has been measured in the range $0 \le \var_m \le 40$ MeV and
$-180 \le p_m \le 270$ MeV/c. The reduced cross sections
for removal of nucleons from $1p_{1/2}$ and $1p_{3/2}$ shells in
${}^{16}$O$(e,e'p){}^{15}$N were measured in quasielastic parallel kinematics at
three different beam energies: $\var_i=304, 456$ and 521 MeV. In this
kinematics, $\p_x$ is parallel or antiparallel to $\q$. The data were obtained
by keeping $\vert \p_x \vert$ constant and
changing the transferred momentum, i.e., by varying the scattering angle of the
two detected particles.
The total kinetic energy in the center-of-mass system between the outgoing
proton and the recoiling ${}^{15}$N nucleus was kept constant at 90 MeV.
In Table II of Ref.~\cite{NIK2} are listed the relevant kinematic parameters of
the experiment. The missing momentum is positive for
$\vert \q \vert < \vert \p_x \vert$.

In electron scattering experiments, the lepton and hadron spectrometers are set
in-plane, and therefore the out-of-plane angles are fixed to $\phi_e=0^{\circ}$
and $\phi_p=180^{\circ}$.
The sixfold exclusive cross section can be written as~\cite{BAV5} 
\begin{equation}
  \frac{d^6\sigma}{d\var_{\mu} d\Omega_{\mu} dT_p d \Omega_p} =
  \frac{N_{1\mu 1p}}{N_{tot}}\frac{\sigma_{tot}(\var_{\nu})}
       {\Delta \var_{\mu} \Delta T_p \Delta \Omega_{\mu} \Delta \Omega_p},  
\end{equation}
where $N_{tot}$ is the total number of generated CCQE neutrino events with a
total CCQE cross section $\sigma_{tot}(\var_{\nu})$ at energy $\var_{\nu}$.
$N_{1\mu 1p}$ is the number of selected $1\mu 1p$ events in the differential
phase-space volume bin
$\Delta V = \Delta \var_{\mu} \Delta T_p \Delta \Omega_{\mu} \Delta \Omega_p$ with
the central values
$\var_{\mu},T_p,\Omega_{\mu}, \Omega_p$ and with size of differential bins 
 $\Delta \var_{\mu}, \Delta T_p, \Delta \theta_{\mu}, \Delta \theta_p,
\Delta \phi_{\mu}, \Delta \phi_p$.
The central values and sizes of the differential bins of kinematic variables
are given in Tables I and II, respectively, for every dataset that is
analyzed in this work.
\begin{table}[t]
        \def\arraystretch{1}
         \begin{tabular}{|c| c c c c c c c |}

                \hline \hline
 \multirow{1}{*}{data set} & $\var_{i}$ & $\var_{f}$ & $\theta_{e}$ & T$_{p}$ &
  $\theta_{p}$ & $\Delta\var_{m}$ & notes\\ 
                & (MeV) & (MeV) & (deg) & (MeV) & (deg) & (MeV) & \\
                \hline \hline
   \multirow{1}{*}{Saclay \cite{Saclay}} & \multirow{1}{*}{500} &
   \multirow{1}{*}{372} & \multirow{1}{*}{59} & \multirow{1}{*}{102} &
   \multirow{1}{*}{35 - 88} & \multirow{1}{*}{10} & \multirow{1}{*}{perpendicular; 1s is not available} \\
   \multirow{1}{*}{NIKHEF \cite{NIK1,NIK2}} & \multirow{1}{*}{304-521} &
   \multirow{1}{*}{188-406} & \multirow{1}{*}{29-81} & \multirow{1}{*}{89-99} &
   \multirow{1}{*}{36 - 49} & \multirow{1}{*}{40} &\multirow{1}{*}{parallel; 1s
   is not available}\\  
                \hline \hline
        \end{tabular}
        \def\arraystretch{1.0}
        \caption{Summary of data for the $^{16}$O$(e,e^{\prime}p)$ reaction.
          $\var_{i}$ is the beam energy, $\var_{f}$ is the central electron
          energy, $\theta_{e}$ is the central electron angle, T$_{p}$ is the
          central proton kinetic energy, $\theta_{p}$ is the central proton
          angle, and $\Delta\var_m$ is the range of the missing energy.}
        \label{tab:kinematics1}
\end{table}

\begin{table}[t]
        \def\arraystretch{1.0}
        \begin{tabular}{|c|c c c c c c|}

                \hline\hline
                \multirow{2}{*}{data set} & $\Delta\varepsilon_{f}$ &
                $\Delta\theta_{e}$ & $\Delta$T$_{p}$ & $\Delta\theta_{p}$ &
                $\Delta\phi_{e}$ & $\Delta\phi_{p}$ \\
                & (\%) & (deg) & (\%) & (deg) & (deg) & (deg)\\
                \hline \hline

         \multirow{1}{*}{Saclay \cite{Saclay}} & \multirow{1}{*}{[-7, +25]} &
         \multirow{1}{*}{2} & \multirow{1}{*}{8} & \multirow{1}{*}{2} &
         \multirow{1}{*}{6} & \multirow{1}{*}{6} \\

         \multirow{1}{*}{NIKHEF \cite{NIK1,NIK2}} & \multirow{1}{*}{[-4, +6]} &
         \multirow{1}{*}{4} & \multirow{1}{*}{[-5.4, +5]} & \multirow{1}{*}{6} &
         \multirow{1}{*}{6} & \multirow{1}{*}{6} \\
            \hline \hline
            
        \end{tabular}
        \def\arraystretch{1.0}
        \caption{Table of the experimental cuts that are used in simulation.
          $\Delta\varepsilon_{f}$ is the electron energy acceptance,
          $\Delta\theta_{e}$ is the electron angle acceptance,
          $\Delta$T$_{p}$ is the proton kinetic energy acceptance, 
        $\Delta\phi_{p}$ is the proton angle acceptance, and $\Delta\phi_{e}$ is
          the electron plane angle acceptance.}
        \label{tab:kinematics2}
\end{table}
To transform from measured variables $\var_{\mu}$ and $T_p$ to phase-space
($\var_m, p_m$)~\cite{BAV5}
\begin{equation}
 \var_m = \omega -T_p - \var_B                               
\end{equation}
and
\begin{equation}
  p_m = [p^2_x + k^2_{\nu} + k^2_{\mu} -2k_{\mu} k_{\nu}\cos\theta_{\mu} -
    2p_xk_{\nu}\cos\theta_p + 2k_{\mu}p_x\cos\theta_{\mu p}]^{1/2}        
\end{equation}
we use the formula Ref.~\cite{Frulani}
\begin{equation}
  \label{pm}
 \frac{d^6\sigma}{dp_m d\var_m d\Omega_{\mu}d \Omega_p}=
  \frac{d^6\sigma}{d\var_{\mu} dT_p d\Omega_{\mu} d \Omega_p}/J(\theta_p),  
\end{equation}
where the Jacobinan $J=\partial(\var_m,p_m)/\partial(\var_{\mu},T_p)$ is equal to
~\cite{BAV5}
\begin{equation}
J(\theta_p)=\frac{1}{p_m}(k_{\mu} - k_{\nu}\cos\theta_{\mu} + p_x\cos\theta_{\mu p})
  - \frac{\var_x}{2 p_m}\frac{p^2_x + p^2_m -|\q|^2}{p^2_x}.       
\end{equation}
and is determined at the bin centers of the kinematic variables.
In Eq. (\ref{pm}), $\cos\theta_{\mu p}=\cos(\theta_{\mu} + \theta_p)$, and we
assess
positive (negative) values to $p_m$ for the condition $\theta_p <\theta_q
(\theta_p > \theta_q)$, where $\cos\theta_q = \p_x\cdot\q/(|\p_x||\q|)$.
Then, we can determine the distorted spectral function as
\begin{equation}
  S^D(p_m,\var_m)=\frac{d^6\sigma}{dp_m d\var_m d\Omega_{\mu} d\Omega_p}/
  {K^{(el)(cc)}\sigma_{\nu N}},  
\end{equation}
where $K^{(el)(cc)}$ is phase-space factor~\cite{BAV5}.
All nuclear models which are implemented in the GENIE version 3 generator do
not take into account the shell structure of ${}^{16}$O and use the 
nucleon momentum distribution as a function of missing momentum at the fixed
value of missing (binding) energy $\var_m$. Therefore, we can evaluate only
\begin{equation}
 \sigma_{red}(p_m)= \int S^D(p_m,\var_m)d\var_m           
\end{equation}
as a function of $p_m$, integrated over the range $\Delta \var_m$, that is
given in Table I. Then, the reduced cross section can be written as
\begin{equation}
  \sigma^{red}(p_m)=\Delta\var_mS^D(p_m).               
  \label{red}
\end{equation}
We calculate the $\sigma_{\nu N}$ using the same nucleon form
factors as in the GENIE cross section calculation and the CC1 de Forest
prescription~\cite{deForest}.


\subsection{GENIE simulation framework}

The GENIE version 3 simulation framework utilizes several different
models of nucleon momentum distribution in the nuclear ground state. It also
offers several models of quasielastic lepton-nucleus interactions and several
intranuclear cascade models for FSI.
We use  GENIE version 3.4.0 and consider four distinct sets of GENIE
models for CCQE scattering on oxygen (see Table III), namely, the G18\_2a,
G18\_10a, G21\_11 (SuSAv2), and effsf (effective spectral function) models
~\cite{G1, G2, G3}.
These models are accompanied by the FSI models hA2018
and hN2018 available in GENIE~\cite{G1} and the hA2018 model is used for each
of our four considered CCQE scattering models.
                \begin{table}[t]
                        \def\arraystretch{1.2}
                        \begin{tabular}{|c|c|}
                                \hline
                                GENIE configuration & Nuclear model \\
                                \hline
                                G18\_02a & Relativistic Fermi Gas \\
                                \hline
                                G18\_10a & Local Fermi Gas \\
                                \hline
                                G21\_11v2 & Local Fermi Gas \\
                                \hline
                                effsf & Effective Spectral Function \\
                                \hline
                        \end{tabular}
       \caption{Table of GENIE configurations and models of momentum
        distribution of nucleons, corresponding to these configurations.}
        \label{tab:configuration}
                \end{table}
\begin{figure*}
  \begin{center}
    \includegraphics[height=8cm, width=15cm]{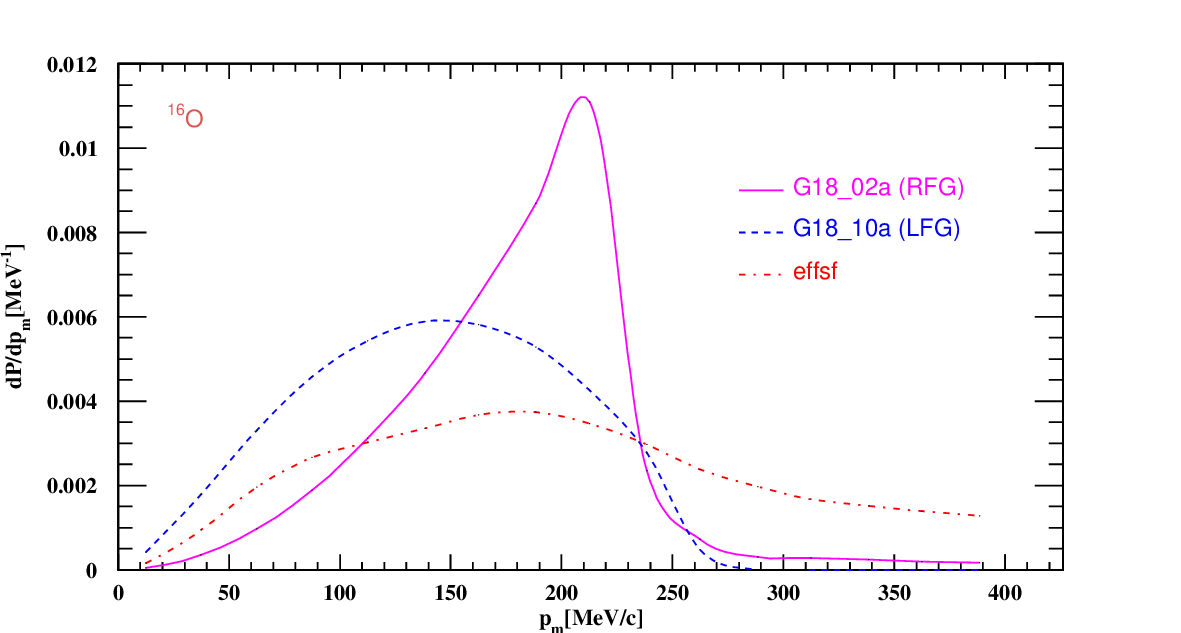}
  \end{center}
  \caption{\label{Fig1}
Initial nucleon momentum distribution for ${}^{16}$O according to the GENIE
implementation of G18\_02a (FG, solid line), G18\_10a (LFG, dashed line), and 
effsf (dot-dashed) models.
}
\end{figure*}

\begin{figure*}
  \begin{center}
    \includegraphics[height=9cm, width=18cm]{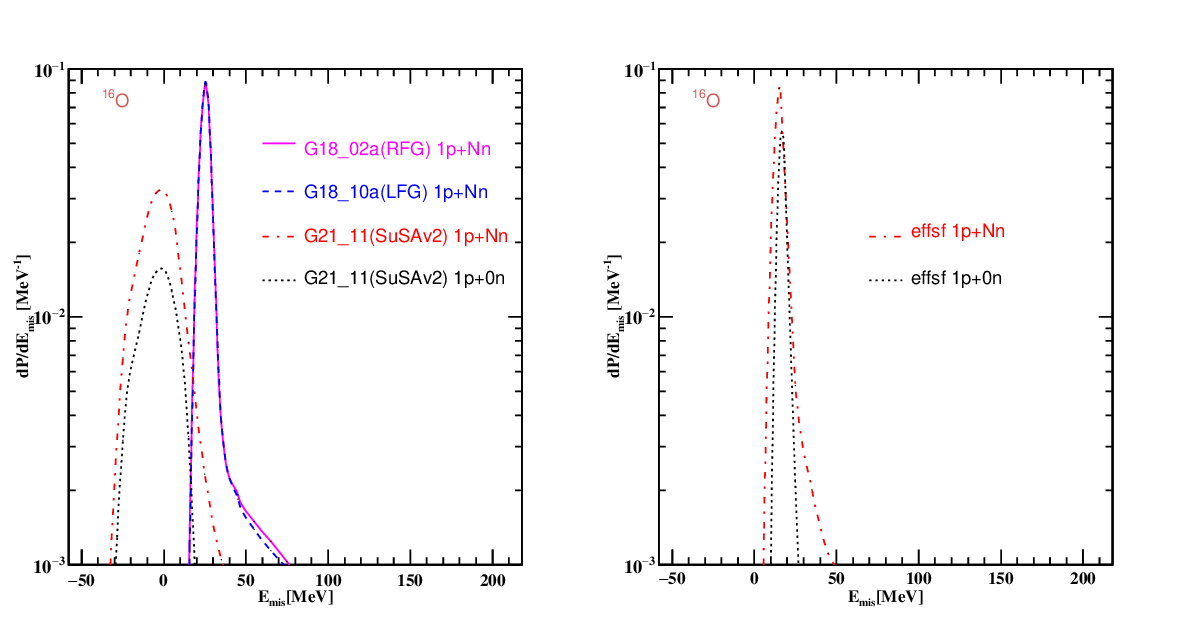}
  \end{center}
  \caption{\label{Fig2}
    Probability density distribution vs missing energy $E_{mis}$ calculated for
    neutrino energy $\var_{\nu}=0.5$ GeV. Left panel: PDF for (1p+Nn) events
    calculated with the G18\_02a (solid line), G18\_10a (dashed line), and
    G21\_11 (dot-dashed line) models and for (1p+0n) events calculated within
    the G21\_11 (dotted line) model.
 Right panel: PDF for the (1p+0n) set calculated with the effsf (dotted line)
model and PDF for the (1p+Nn) set calculated within the effsf (dot-dashed line)
 models. 
}
\end{figure*}

The G18\_02a configuration of the GENIE 
relies on implementation of the relativistic Fermi gas model (RFG), which has
been modified to incorporate short-range nucleon-nucleon correlations~
\cite{BodR}. The CCQE scattering is simulated by the Llewellyn-Smith model
~\cite{LS}. 
The G18\_10a model set includes the full Valencia model~\cite{Val1, Val2, Val3}
for the local Fermi gas (LFG) nucleon momentum distribution. It also utilizes 
the random phase approximation (RPA) which is a description of long-range and
short-range NN correlations.

The G21\_11 (SuSAv2) model is based on a superscaling model that accurately
describes QE inclusive scattering on a variety of targets (A $\ge$ 10) for a
wide range of electron energies. The version GENIE v3 describes
struck nucleon momenta with a mean-field model (similar to LFG) at low
momentum transfer and with a relativistic Fermi gas at high momentum transfer.
To describe the outgoing nucleon kinematics the LFG nuclear model is used.
The effective spectral function (effsf) model with or without enhancement of
the transverse contribution~\cite{Bod1, Bod2} is implemented in GENIE as the
option EffectiveSF.

The bound nucleon momentum distributions in oxygen for the
genuine CCQE events produced using the GENIE and the RFG, LFG, and effsf
representations of the target nucleus are shown in Fig.\ref{Fig1}. It is clearly
visible that in the range $150 \le p_m \le 220$ ($p_m \le 50$) (MeV/c) the
probability density function (PDF) $dP/dp_m$ calculated with the RFG is much
larger (smaller) than the PDFs used in the other models, and correlation tails
are observed in the RFG and effsf distributions. The missing energy
$E_{mis}=\var_{\nu} - \var_{\mu} - T_p$ distribution for two sets of events in
which neutrino knocks out a proton, calculated with neutrino energy
$\var_{\nu}=0.5$ GeV using four GENIE models is shown in Fig.\ref{Fig2}. The
first set contains events with a muon and only one proton (1p + 0n) in the
final state.
\begin{figure*}
  \begin{center}
    \includegraphics[height=11cm, width=18cm]{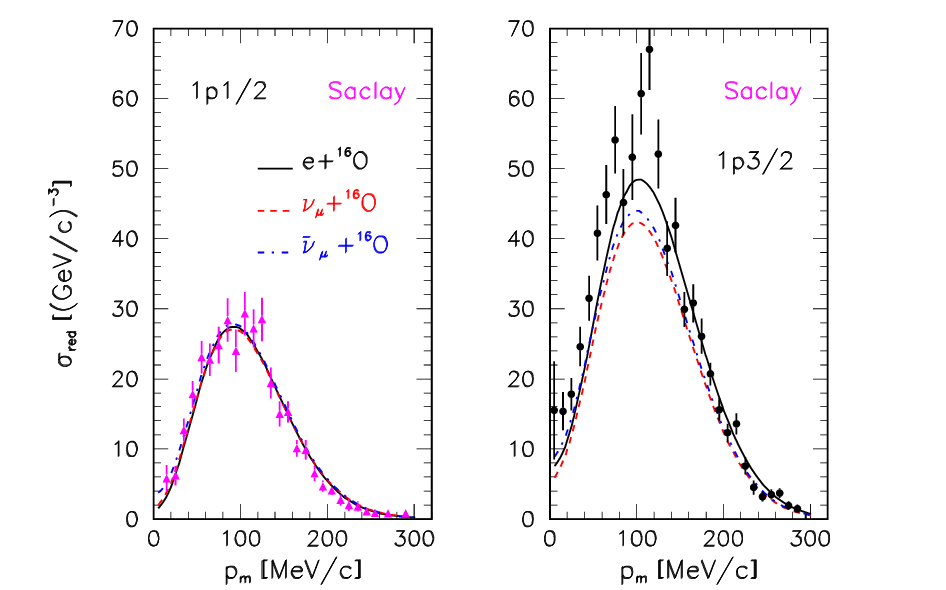}
  \end{center}
  \caption{\label{Fig3}
    Comparison of the RDWIA electron, neutrino, and antineutrino reduced cross
    sections~\cite{BAV1} for the removal of nucleons from $1p_{1/2}$ and
    $1p_{3/2}$ shells of oxygen for Saclay~\cite{Saclay} perpendicular
    kinematics. Saclay data for beam energy $E_{beam}=500$  MeV, proton kinetic
    energy $T_p=100$ MeV and $Q^2=0.3$GeV${}^2$.
}
\end{figure*}
Such events are selected in electron scattering experiments
$(e,e'p)$, where events with one or more low energy neutrons are only excluded
through kinematic cuts. The events from (1p + 0n) set are used to measure
reduced cross sections.
The second set contains events
with a muon, one proton, and at least one neutron (1p + Nn) in the final state.
This set also includes the events in which a knocked out protons produce
neutrons in inelastic interactions with the residual nucleus, i.e. events with
large missing energy.
In neutrino experiments these two sets of events
are not distinguished, because the incident neutrino energy is unknown. For all
models (except G21\_11 model), the maximum of distribution is located in the
range $E_m \le 40$ MeV. For the G18\_02a and G18\_10a models, the distributions
of events in the (1p+0n) set have narrow peak in the range $20 \le E_m \le 30$
MeV. The inelastic scattering changes the energy of the knocked-out protons
significantly, and therefore the distributions of events in the (1p+Nn) set are
extend into the higher missing region. 
\section{Results and analysis}

In Saclay and NIKHEF experiments the reduced cross sections were measured as a
\begin{figure*}
  \begin{center}
    \includegraphics[height=14cm, width=14cm]{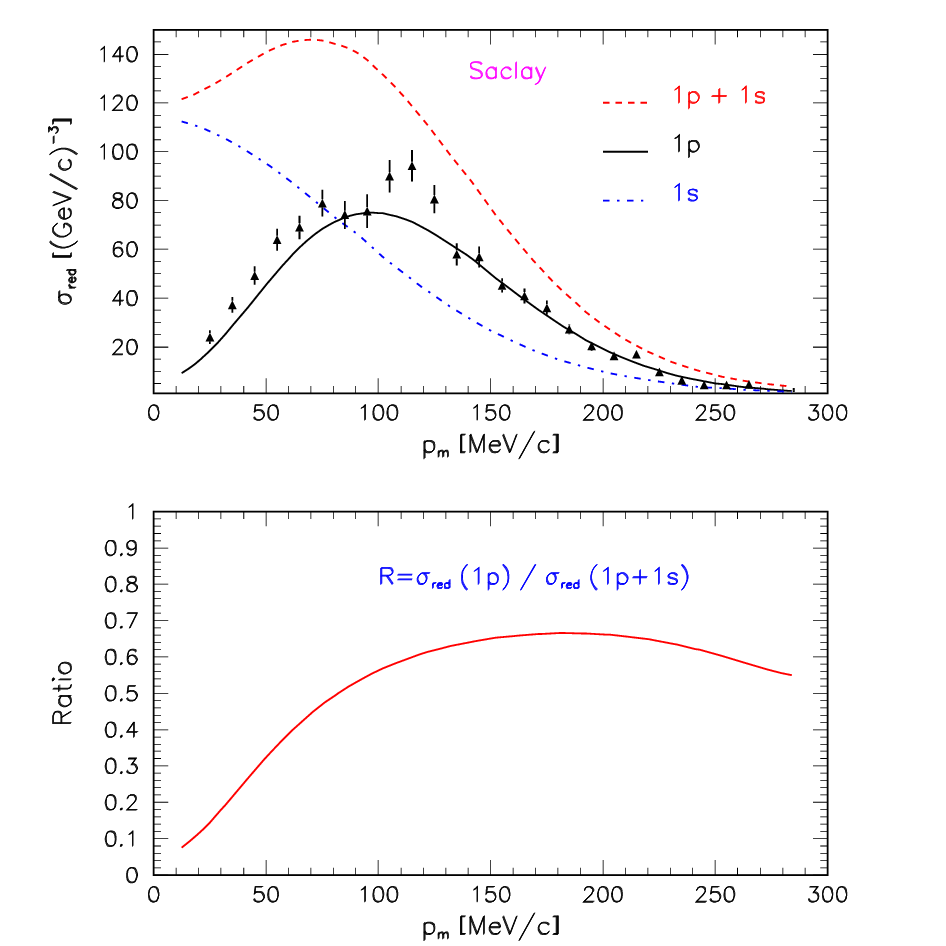}
  \end{center}
  \caption{\label{Fig4}
 The RDWIA electron reduced cross section~\cite{BAV1} for the removal of protons
 from $1s$, $1p$, and $1s+1p$ shells of ${}^{16}$O for Saclay kinematics (upper
 panel) and ratio $R_{1p}$ (lower panel) as functions of missing momentum. Also
 shown (upper panel) are Saclay data for the removal of protons from
 $1p=1p_{1/2}+1p_{3/2}$ shell.      
}
\end{figure*}
function of missing momentum for the removal of a proton from $1p_{1/2}$ and
$1p_{3/2}$ shells of ${}^{16}$O which correspond to the intervals of missing
energy $\var_m=$10-15 MeV and 15-20 MeV, respectively.
\begin{figure*}
  \begin{center}
    \includegraphics[height=11cm, width=18cm]{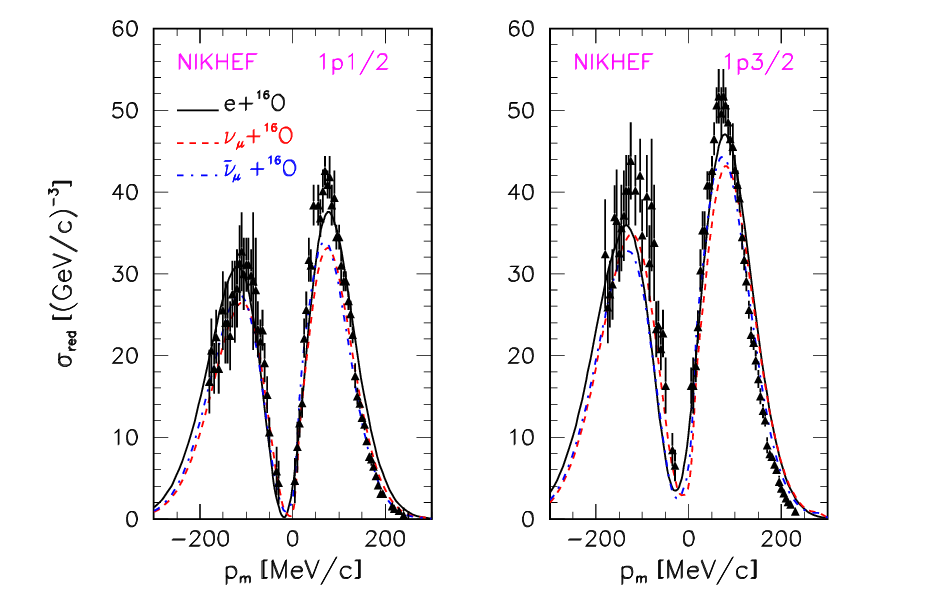}
  \end{center}
  \caption{\label{Fig5}
    The RDWIA calculation of electron, neutrino, and antineutrino reduced cross
    sections~\cite{BAV1} for the removal of nucleons from $1p_{1/2}$ and
    $1p_{3/2}$ shells of ${}^{16}$O for NIKHEF parallel kinematics~
    \cite{NIK1,NIK2} as function of $p_m$. NIKHER data for beam energy
    $E_{beam}=304-521$ MeV, proton kinetic energy $T_p=89-99$ MeV and
    $Q^2$ is varied.    
}
\end{figure*}
The models which are used in the GENIE generator to calculate the reduced
cross sections do not take into account the shell structure of the ${}^{16}$O
nucleus. Therefore we cannot calculate the reduced cross sections for the
removal of nucleons separately from the $1s$ and $1p$ shells.

The reduced exclusive cross sections for the removal of nucleons from $1p_{1/2}$
and $1p_{3/2}$ shells in ${}^{16}$O$(e,e'p){}^{15}$N,
${}^{16}$O$(\nu,\mu^- p){}^{15}$O, and ${}^{16}$O$(\bar{\nu},\mu^+ n){}^{15}$N
reactions calculated within the RDWIA~\cite{BAV1} are shown in Fig.~\ref{Fig3}
together with Saclay data.
The cross sections were calculated using the Saclay
kinematic conditions with the normalization factors of data
examined presented in Ref.~\cite{Fissum}.
The small difference between neutrino and antineutrino reduced cross sections
for the removal of nucleons from $1p_{1/2}$ and $1p_{3/2}$ states is due to
different nucleon's binding energies, that are higher for neutrons, and nucleon
momentum distributions for these shells. The FSI effects are also different
between the cross sections due to interaction of the outgoing nucleons with
different residual nuclei $p+{}^{15}$O for neutrino and $n+{}^{15}$N for
antineutrino scattering off ${}^{16}$O and Coulomb interaction of the knocked
out protons. Moreover, the $1p_{1/2}$ orbit has spatial characteristics which
are different from the $1p_{3/2}$ orbit. The RDWIA model predicts a stronger
attenuation for proton emission from a level which has a larger fraction of
its density in the nuclear interior~\cite{BAV5}. 
The reduced cross sections for the removal of protons from the 
$1p=1p_{1/2}+1p_{3/2}$ shell in ${}^{16}$O$(e,e'p){}^{15}$N reaction together with
Saclay data are shown in Fig.~\ref{Fig4}. Also shown are the reduced cross
section calculated in the perpendicular Saclay kinematics for the knock out
of protons from 1s ($\sigma_{red}(1s)$) and 1s+1p ($\sigma_{red}(1p+1s)$) shells
in the ${}^{16}$O$(e,e'p){}^{15}$N reaction, and a ratio
$R_{1p}=\sigma_{red}(1p)/\sigma_{red}(1p+1s)$ as functions of missing momentum
$p_m$.
\begin{figure*}
  \begin{center}
    \includegraphics[height=14cm, width=14cm]{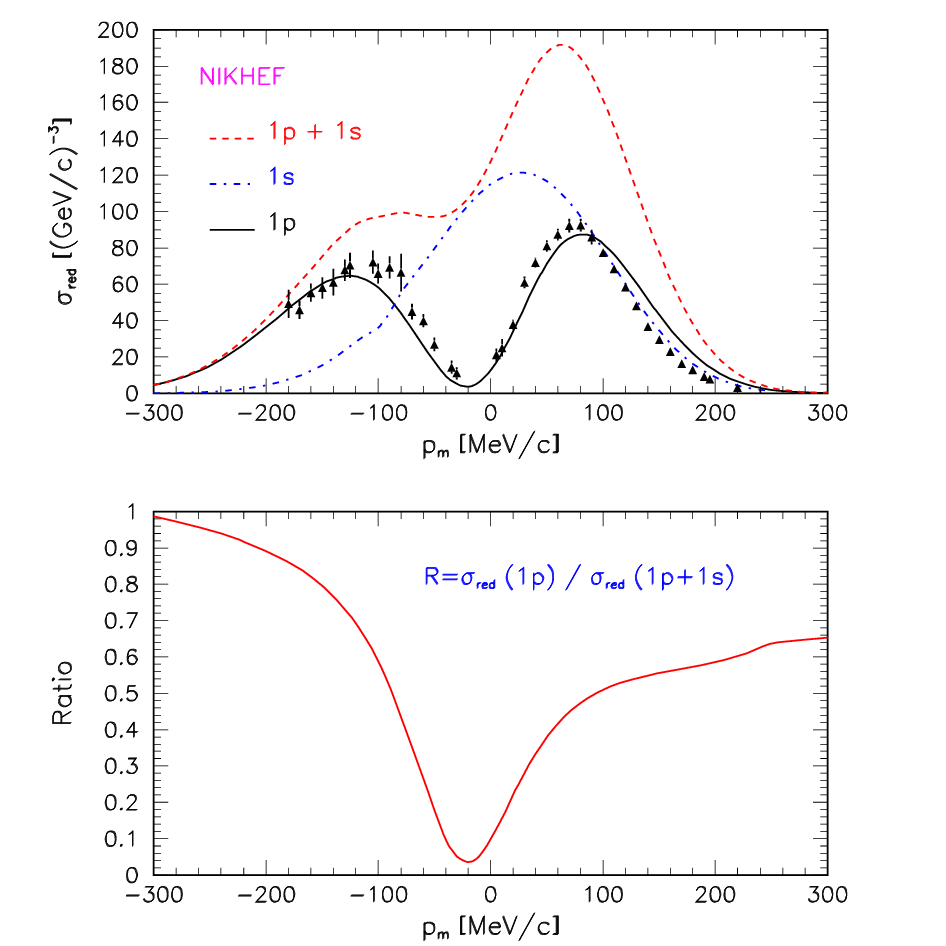}
  \end{center}
  \caption{\label{Fig6}
 The RDWIA electron reduced cross section~\cite{BAV1} for the removal of protons
 from $1s$, $1p$, and $1s+1p$ shells of ${}^{16}$O for NIKHEF kinematics (upper
 panel) and ratio $R_{1p}$ (lower panel) are shown as functions of missing
 momentum. Also shown (upper panel) are Saclay data for the removal of protons
 from $1p=1p_{1/2}+1p_{3/2}$ shell.          
}
\end{figure*}
The contribution of the 1p shell to the reduced cross section calculated
 with the GENIE models can be evaluated using this ratio as   
 \begin{equation}
  \sigma^{MC}_{red}(1p)=R_{1p}\cdot \sigma^{MC}_{red}(1p+1s),               
\end{equation}
 where $\sigma^{MC}_{red}(1p+1s)$ is determined by Eq.(\ref{red}). In the
 following, we compare $\sigma^{MC}_{red}(1p)$ with the measured reduced cross
 sections.
\begin{figure*}
  \begin{center}
    \includegraphics[height=10cm, width=19cm]{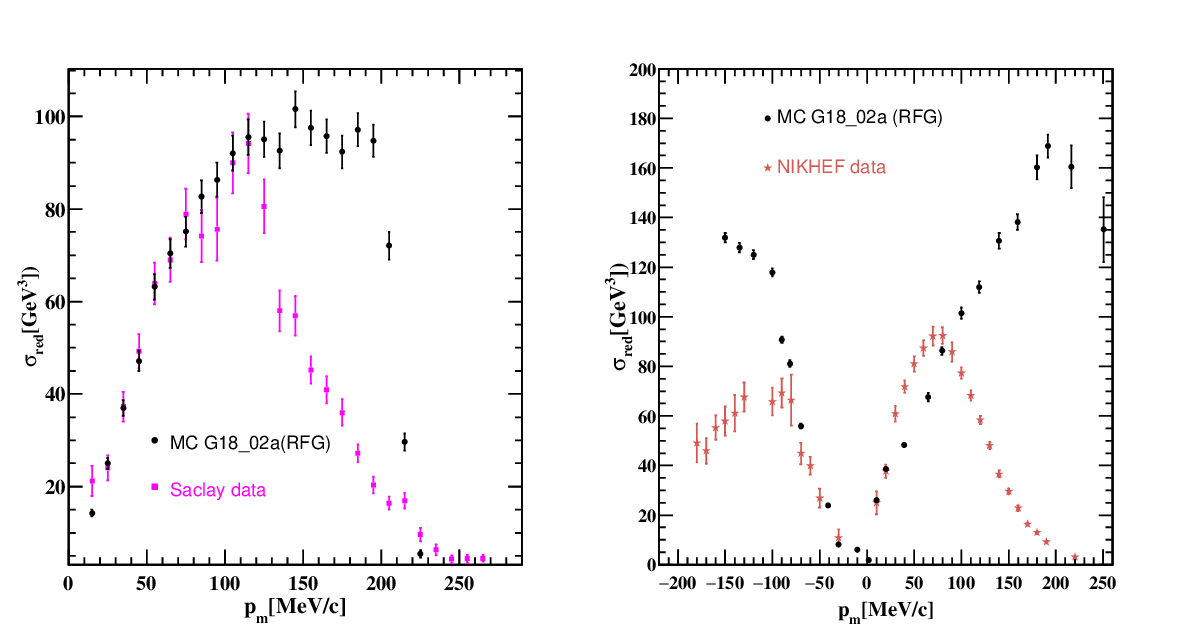}
  \end{center}
  \caption{\label{Fig7}
    Comparison of the GENIE G18\_02a (RFG) model calculation reduced cross
    sections with data as functions of missing momentum. As shown in key,
    cross sections were calculated for Saclay~\cite{Saclay} and
    NIKHEF~\cite{NIK1, NIK2} kinematics.
} 
\end{figure*}
 
 The reduced cross sections together with NIKHEF data are shown in
 Fig.~\ref{Fig5}. The cross sections were calculated using the normalization
 factors for the data examined in Ref.~\cite{Fissum}. The reduced cross
 sections for the knock out of protons from the $1p=1p_{1/2}+1p_{3/2}$ shell
 calculated in the parallel NIKHEF kinematics together with data are shown in
 Fig.~\ref{Fig6}. Also shown in this figure is the ratio $R_{1p}$ as a function
 of missing momentum. This ratio is minimal at $p_m=0$, where the contribution
 to the reduced cross section by removal nucleons from 1s shell is dominant
 and increases with missing momentum up to 0.6-0.7 at $p_m=150$ MeV/c.
 This range corresponds to the knock out of nucleons from 1p shell where the
 reduced cross sections measured at Saclay and NIKHEF kinematics reach maximum
 value of about 80(GeV/c)$^3$. There is an overall good agreement between
 calculated cross sections, but the value of electron cross sections at maximum
 is systematically higher (less than 10\%) than (anti)neutrino ones with the
 exception of the $1p_{1/2}$ state for Saclay kinematics.
\begin{figure*}
  \begin{center}
    \includegraphics[height=10cm, width=19cm]{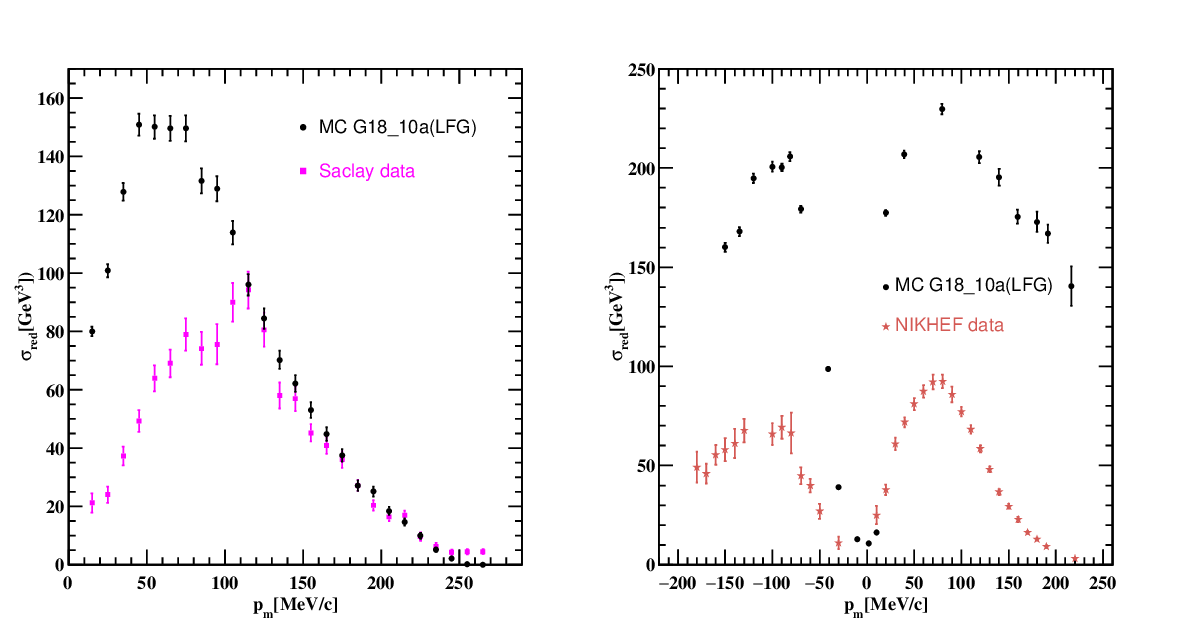}
  \end{center}
  \caption{\label{Fig8}
    Same as Fig.~\ref{Fig7} but for the GENIE G18\_10a (LFG) model
    calculation.
}
\end{figure*}
The small difference between neutrino and antineutrino reduced cross sections
is due to the difference in the FSI of protons and neutrons with residual
nucleus. 
 
For each model considered we generated $10^8$ CCQE neutrino events. The GENIE
results are presented only with statistical errors.
The reduced cross sections calculated with the GENIE G18\_02a
model that uses the RFG nucleon momentum distribution are shown in
Fig.~\ref{Fig7} together with Saclay~\cite{Saclay} and NUKHEF~\cite{NIK1, NIK2,
  NIK3} data. The cross sections were calculated using the kinematic conditions
of the data examined that is presented in Tables I and II. At $|p_m| \le 100$
MeV/c the values of the calculated cross sections are in agreement with the
data and at higher missing momentum the calculated $\sigma_{red}$ overestimates
the measured ones significantly. This excess increases with missing momentum
because of the uniform momentum distribution of the Fermi gas model.
In Fig.~\ref{Fig8} the comparison with data of the reduced cross sections
calculated with the GENIE G18\_10a model are presented.
\begin{figure*}
  \begin{center}
    \includegraphics[height=10cm, width=19cm]{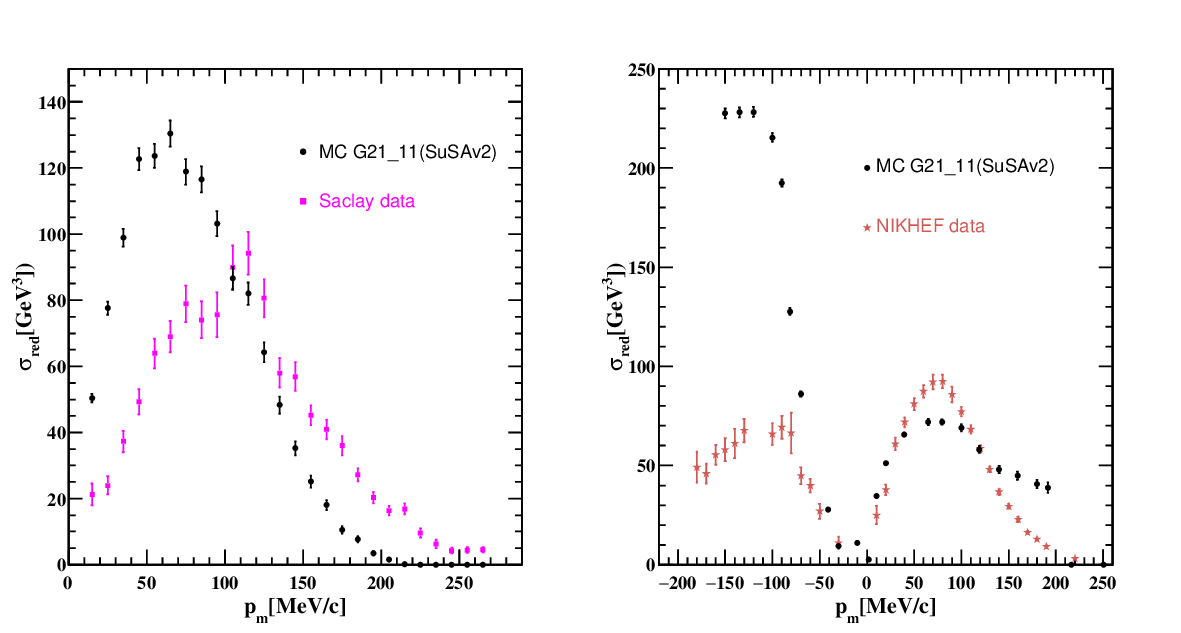}
  \end{center}
  \caption{\label{Fig9}
    Same as Fig.~\ref{Fig7} but for the GENIE G21\_11 (SuSAv2) model
    calculation.
} 
\end{figure*}
We observe agreement between GENIE predictions and Saclay data at
$|p_m| \ge 120$ MeV/c. At low missing momentum the calculation overestimates
data significantly. On the other hand the $\sigma_{red}$ calculated within this
GENIE model for NIKHEF parallel kinematics overestimates the measured electron
scattering data in the range of $|p_m| \ge 50$ MeV/c. At $|p_m| \approx 100$
MeV/c where the maximum of the reduced cross section is observed the excess
reaches up to 300\%.
Figure~\ref{Fig9} shows the reduced cross sections calculated with the GENIE
G21\_11 model where the LFG nucleon momentum distribution is used.
The $\sigma_{red}$ calculated with this model for
Saclay kinematics overestimates significantly measured cross sections at
$p_m \le 100$ MeV/c and underestimates the data in the range of $p_m \ge 150$
MeV/c. At NIKHEF kinematics the issue visible at $p_m \ge 150$ MeV/c resemble
those for the G18\_10a model seen in Fig.~\ref{Fig8}, but the excess is
significant only in the range of $p_m \le -100$ MeV/c.  
\begin{figure*}
  \begin{center}
    \includegraphics[height=10cm, width=19cm]{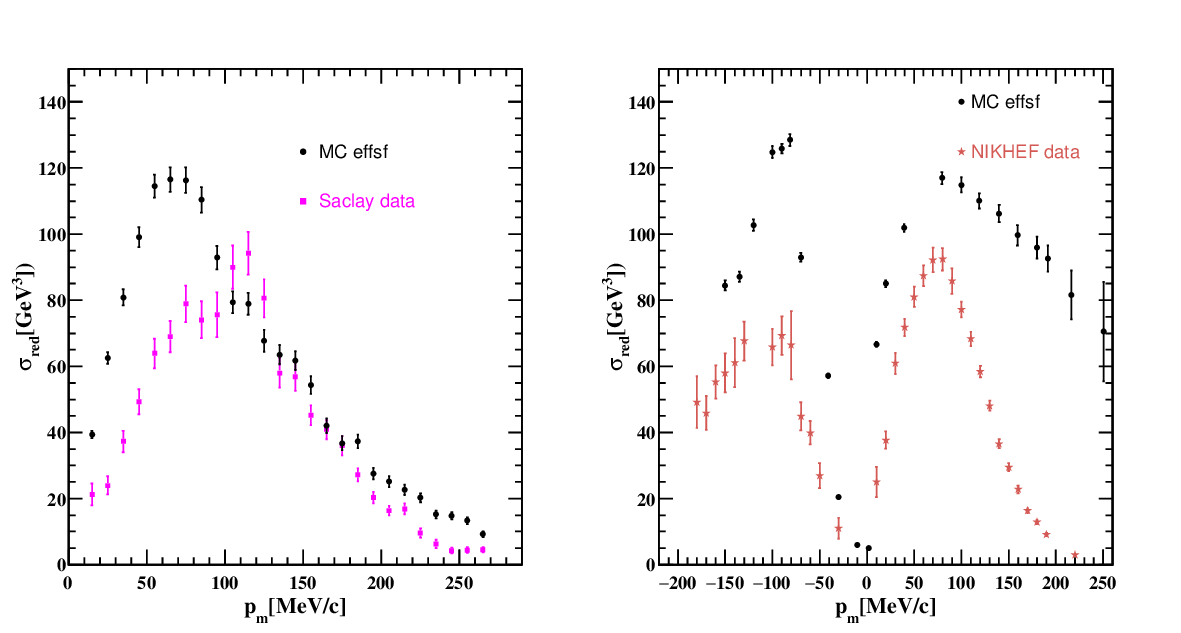}
  \end{center}
  \caption{\label{Fig10}
 Same as Fig.~\ref{Fig7} but for the GENIE effsf model calculation.  
}
\end{figure*}
Approximately the same features are observed in Fig.~\ref{Fig10}, where the
comparison with data of the reduced cross sections calculated with the GENIE
effsf model is presented. The calculated cross sections overestimate the
measured one at Saclay less than 60\% at $p_m \le 100$ MeV/c. The difference
between the reduced cross sections calculated and measured at NIKHEF kinematics
is significant at $|p_m| \ge 100$ MeV/c.

So, at Saclay perpendicular kinematics the agreement of the GENIE calculated
reduced cross sections with data for neutrino scattering off oxygen and
carbon ~\cite{BAV4} is approximately the same. For models considered (except
the RFG), the issues are similar: the
calculations overestimate data significantly at low missing momenta and
underestimate them slightly (except SuSAv2 model) at $p_m \ge 100$ MeV/c. The
$\sigma_{red}$ calculated at NIKHEF parallel kinematics demonstrates absolutely
different behavior (except the RFG model) than measured ones. We observe
persistent disagreement between the GENIE predictions of the
reduced cross sections and electron scattering data at low beam energy.    
On the other hand there is a good agreement between the RDWIA calculated and
measured cross sections and this approach can be used to model both
lepton-boson and boson-nucleus vertices~\cite{NEUT_RD}. It is obvious that
this model describes the semiexclusive $(l,l'p)$ lepton scattering process
better than the phenomenological models employed in the GENIE simulation
framework.  
\section{Conclusions}

In this article, we carried out a systematic comparison of the CCQE reduced
cross sections calculated with models employed in the GENIE version 3
simulation framework with data measured in electrons scattering off the oxygen
target. The reduced
cross sections as functions of missing momentum were measured only for the
removal of protons from $1p$ shells of ${}^{16}$O as a whole, i.e. from the
$1p+1s$ shells. The reduced cross sections for the removal of proton from the
$1p$ shells calculated with the GENIE models were evaluated using the relative
contribution
of the $1p$ shells $R_{1p}$ predicted within the RDWIA model at Saclay and
NIKHEF kinematics. We have observed persistent disagreement between the GENIE
predictions and electron scattering data for the reduced cross sections. The
$\sigma_{red}$ calculated at NIKHEF parallel kinematics demonstrates absolutely
different behavior than the measured ones. At Saclay perpendicular kinematics
the agreement with data of the reduced cross sections calculated within the
GENIE v.3 simulation
framework for neutrino scattering on oxygen and carbon targets is similar: the
calculations overestimate the data at low missing momenta and underestimate the
data at $p_m \ge 100$ MeV/c. Therefore, the GENIE event generator cannot
simulates well two tracks CCQE events in the all allowed kinematic region.
Neutrino event
genetators need to use more sophisticated models to simulate CCQE
semiexclusive processes. The direct comparison of spectral functions,
implemented in neutrino event generators, with the precise electron reduced
cross section data is an original and promising method for testing the quality
of nuclear physics models implemented in Monte Carlo generators for neutrino
interuction. 
\section{Data availability}

The data that support the findings of this article are openly available~\cite
{Saclay, NIK1, NIK2}

\section*{Acknowledgments}

We thank Dr. Lapikas and Dr. Jans for the specifications of spectrometers that
 were used in NIKHEF experiment. 
%


\end{document}